\documentclass{article}

\usepackage{arxiv}

\usepackage[utf8]{inputenc} 
\usepackage[T1]{fontenc}    
\usepackage{hyperref}       
\usepackage{url}            
\usepackage{booktabs}       
\usepackage{amsfonts}       
\usepackage{nicefrac}       
\usepackage{microtype}      
\usepackage{lipsum}
\usepackage{graphicx}
\usepackage{amsmath}
\usepackage{xcolor}
\usepackage{cite}
\usepackage{subcaption}
\newcommand{\etal}{\emph{et al.}}

\graphicspath{ {./images/} }

\title{xLSTM-ECG: Multi-label ECG Classification via Feature Fusion with xLSTM}

\author{Lei Kang\textsuperscript{$\dagger$}, Xuanshuo Fu\textsuperscript{$\dagger$}, Javier Vazquez-Corral, Ernest Valveny, Dimosthenis Karatzas\\[1ex]
Computer Vision Center, Universitat Autònoma de Barcelona, Spain.\\
\{lkang, xuanshuo, javier.vazquez, ernest, dimos\}@cvc.uab.es
}

\begin{document}
\maketitle

\begingroup
\renewcommand{\thefootnote}{}%
\footnotetext{\textsuperscript{\dag}These authors contributed equally to this work.}
\endgroup

\begin{abstract}
Cardiovascular diseases (CVDs) remain the leading cause of mortality worldwide, highlighting the critical need for efficient and accurate diagnostic tools. Electrocardiograms (ECGs) are indispensable in diagnosing various heart conditions; however, their manual interpretation is time-consuming and error-prone. In this paper, we propose xLSTM-ECG, a novel approach that leverages an extended Long Short-Term Memory (xLSTM) network for multi-label classification of ECG signals, using the PTB-XL dataset. To the best of our knowledge, this work represents the first design and application of xLSTM modules specifically adapted for multi-label ECG classification. Our method employs a Short-Time Fourier Transform (STFT) to convert time-series ECG waveforms into the frequency domain, thereby enhancing feature extraction. The xLSTM architecture is specifically tailored to address the complexities of 12-lead ECG recordings by capturing both local and global signal features. Comprehensive experiments on the PTB-XL dataset reveal that our model achieves strong multi-label classification performance, while additional tests on the Georgia 12-Lead dataset underscore its robustness and efficiency. This approach significantly improves ECG classification accuracy, thereby advancing clinical diagnostics and patient care. The code will be publicly available upon acceptance.
\end{abstract}

\section{Introduction}
\label{sec:introduction}

Cardiovascular diseases (CVDs) are the leading cause of mortality worldwide, accounting for more than 30\% of global deaths~\cite{lindstrom2022global}. Electrocardiograms (ECGs) are a primary non-invasive tool for detecting conditions such as myocardial infarction, ischemia, and arrhythmias. However, manual interpretation of ECG signals can be time-consuming and prone to error due to subtle waveform variations. This highlights the need for automated ECG classification methods that can assist healthcare professionals in delivering accurate and timely diagnoses.

Recent advances in automated medical analysis have shown that deep learning methods can be highly effective across various domains. For example, a robust fusion framework presented in~\cite{iqbal2025end} integrates diverse feature representations to identify endothelial cells accurately, while deep CNN architectures have demonstrated high performance in dermatological imaging tasks~\cite{iqbal2021automated}. In the context of ECG analysis, deep learning allows models to learn complex patterns directly from raw signals. Although many studies focus on single-lead ECGs, the limited spatial information in single leads can be insufficient for diagnosing more complex cardiac abnormalities. Twelve-lead ECGs provide a more comprehensive view of the heart’s electrical activity from multiple spatial angles, enhancing diagnostic accuracy. Nevertheless, effectively automating the analysis of 12-lead signals remains challenging due to higher data complexity and the necessity of modeling both local and global patterns.

Convolutional Neural Networks (CNNs) naturally excel at recognizing local features via small, fixed-size kernels~\cite{Bai2018AnEmpirical,LeCun2015DeepLearning}, yet they lack a clear mechanism for handling longer-term dependencies~\cite{Schirrmeister2017DeepLearning}. Although deeper CNN architectures can partially expand the receptive field, vanishing gradients~\cite{Glorot2010Understanding} and increased complexity often impede their ability to capture extended temporal relationships. In contrast, Recurrent Neural Networks (RNNs), Long Short-Term Memory (LSTM) networks, and Transformers incorporate dedicated mechanisms—such as memory cells and attention—to better retain and model long-term dependencies~\cite{Hochreiter1997LongShortTerm,vaswani2017attention}.

The 12-lead ECG signals are more complex than single-channel data because each lead captures distinct spatial information~\cite{Kligfield2007Recommendations}. Standard LSTMs, designed primarily for one-dimensional sequences, may not adequately leverage inter-lead correlations in 12-lead ECGs, where channel interactions often carry critical diagnostic clues~\cite{Strodthoff2021DeepLearning,Tan2021MultiLead}. While improved feature-fusion methods or cross-channel attention mechanisms can better capture these relationships~\cite{Xiong2021ECGAttention}, the presence of noise and artifacts across different leads makes robust modeling particularly challenging. Consequently, there is a pressing need for architectures capable of addressing both temporal and spatial complexities.

To tackle these issues, we propose a novel multi-label classification method for 12-lead ECG signals based on the Extended Long Short-Term Memory (xLSTM)~\cite{beck2024xlstm} architecture, which expands the memory capacity beyond standard LSTMs and is well-suited to capturing the intricate patterns in ECG data. Specifically, xLSTM integrates scalar (sLSTM) and matrix (mLSTM) modules to handle both temporal dynamics within each lead and inter-lead dependencies more effectively. We further enhance the model by combining xLSTM with Short-Time Fourier Transform (STFT) preprocessing, transforming time-domain signals into the frequency domain to enrich the overall feature representation.

Although xLSTM has been introduced relatively recently, it has not been fully exploited for 12-lead ECG analysis. In this work, we adapt xLSTM with a specialized dual-module design and a novel feature fusion strategy tailored to handle both local and global patterns, as well as the unique noise characteristics found in multi-lead recordings. Experimental results on the large-scale PTB-XL dataset~\cite{wagner2020ptb} and Georgia 12-Lead dataset~\cite{alday2020classification} show that our xLSTM-based approach substantially outperforms standard LSTM architectures, offering a promising solution for accurate and efficient multi-lead ECG classification.

Our main contributions are summarized as follows: 

\begin{itemize} 
    \item We present a custom-designed xLSTM-based network for multi-label classification of 12-lead ECG signals. This is, to our knowledge, the first application of xLSTM networks in this context. 
    \item We employ Short-Time Fourier Transform to convert ECG signals into the frequency domain, thereby capturing both temporal and spectral characteristics that aid learning. 
    \item We thoroughly evaluate our approach on the PTB-XL dataset, demonstrating state-of-the-art performance, and further assess its robustness on the Georgia 12-Lead dataset, confirming its potential for clinical application.
\end{itemize}

\section{Related Work}

Early ECG classification methods often relied on manual or semi-automatic feature extraction followed by classical machine learning algorithms. For instance, Nasiri~\etal~\cite{nasiri2009ecg} extracted 22 time-voltage features and then utilized Support Vector Machines (SVM)\cite{hearst1998support}, while Gutiérrez-Gnecchi\etal~\cite{gutierrez2017dsp} applied wavelet transform for feature extraction before employing fully connected layers for classification.

With the growing success of deep learning, end-to-end approaches that learn directly from raw ECG data have gained traction~\cite{hannun2019cardiologist}. Recent deep learning methods can be broadly categorized into four main groups: Convolutional Neural Networks (CNN)\cite{lecun2015deep}, Recurrent Neural Networks (RNN)\cite{hopfield1982neural}, Graph Neural Networks (GNN)\cite{scarselli2008graph}, and Transformers\cite{vaswani2017attention}.

\textbf{CNN-based approaches.} Zhang~\etal~\cite{zhang2020se} proposed SE-ECGNet, a multi-scale deep residual network using both 2D and 1D convolutions alongside the Squeeze-and-Excitation module to boost classification accuracy. Jing~\etal~\cite{jing2023ecg} introduced a Beat-Level Fusion Network (BLF-Net) that utilizes an attention mechanism to weight individual heartbeats, thereby improving diagnostic precision.

\textbf{RNN-based approaches.} Murugesan~\etal~\cite{murugesan2018ecgnet} combined CNN and BLSTM modules to exploit hierarchical information at the beat, rhythm, and channel levels, with attention-based visualizations for interpretability. Mousavi~\etal~\cite{mousavi2019inter} employed a sequence-to-sequence model for robust intra- and inter-patient heartbeat classification. Martin~\etal~\cite{martin2021real} showcased an LSTM network that achieved state-of-the-art accuracy across multiple datasets and sampling frequencies, enabling real-time deployment on portable devices for myocardial infarction detection.

\textbf{GNN-based approaches.} Zhao~\etal~\cite{zhao2022ecgnn} introduced ECGNN, which incorporates a feature extractor and a GNN module to capture inter-lead relationships in ECG signals. Ge~\etal~\cite{ge2024knowledge} proposed ECG-KG, leveraging knowledge graph constructions of abnormal ECGs to enhance classification.

\textbf{Transformers-based approaches.} Zhou~\etal~\cite{zhou2023masked} utilized masked autoencoders for self-supervised learning of ECG time-series representations, implementing a lightweight Transformer encoder and a single-layer Transformer decoder.

Beyond these main categories, multiple techniques have been proposed to refine model performance. Golany~\etal~\cite{golany2021ecg} incorporated physical insights by learning ordinary differential equations that describe ECG dynamics, while Rawi~\etal~\cite{rawi2023deep} employed the Tree of Parzen Estimator (TPE)\cite{bergstra2011algorithms} for hyperparameter tuning. Qin\etal~\cite{qin2023mvkt} proposed a teacher-student framework in which a multi-lead ECG teacher model guides a student network trained on single-lead data, leveraging multi-label disease knowledge distillation. Liu~\etal~\cite{liu2024etp} introduced ECG-Text Pre-training (ETP), a cross-modal method that combines ECG and text data for robust classification and zero-shot tasks. Li~\etal~\cite{li2024frozen} developed a Multimodal ECG-Text Self-supervised (METS) approach that learns joint ECG-text representations for zero-shot classification.

CNN-based methods using multi-scale convolutions or 2D representations have shown promise in capturing localized patterns but often struggle with long-range dependencies and inter-lead correlations that are crucial for distinguishing complex cardiac conditions. RNNs and LSTMs can model sequential data but commonly treat each lead independently, overlooking spatial relationships critical in 12-lead ECGs. Meanwhile, GNNs and Transformers, though effective in capturing global interactions, can be computationally heavy and require extensive parameter tuning, limiting their applicability in real-time scenarios.

Extended Long Short-Term Memory (xLSTM)\cite{beck2024xlstm} has recently emerged as a robust enhancement over traditional LSTM architectures, featuring state expansion and advanced gating, normalization, and stabilization mechanisms. Studies suggest that xLSTM can outperform standard LSTMs and even some Transformer models in complex temporal modeling tasks\cite{schneider2024comes}. However, its potential in multi-lead ECG analysis remains largely unexplored. The present work aims to fill this gap by adapting xLSTM with a dual-module design and a specialized fusion strategy to capture the intricate spatial and temporal features required for accurate 12-lead ECG classification.

\section{xLSTM-ECG Method}

\begin{figure*}
  \centering
  \includegraphics[width=0.93\linewidth]{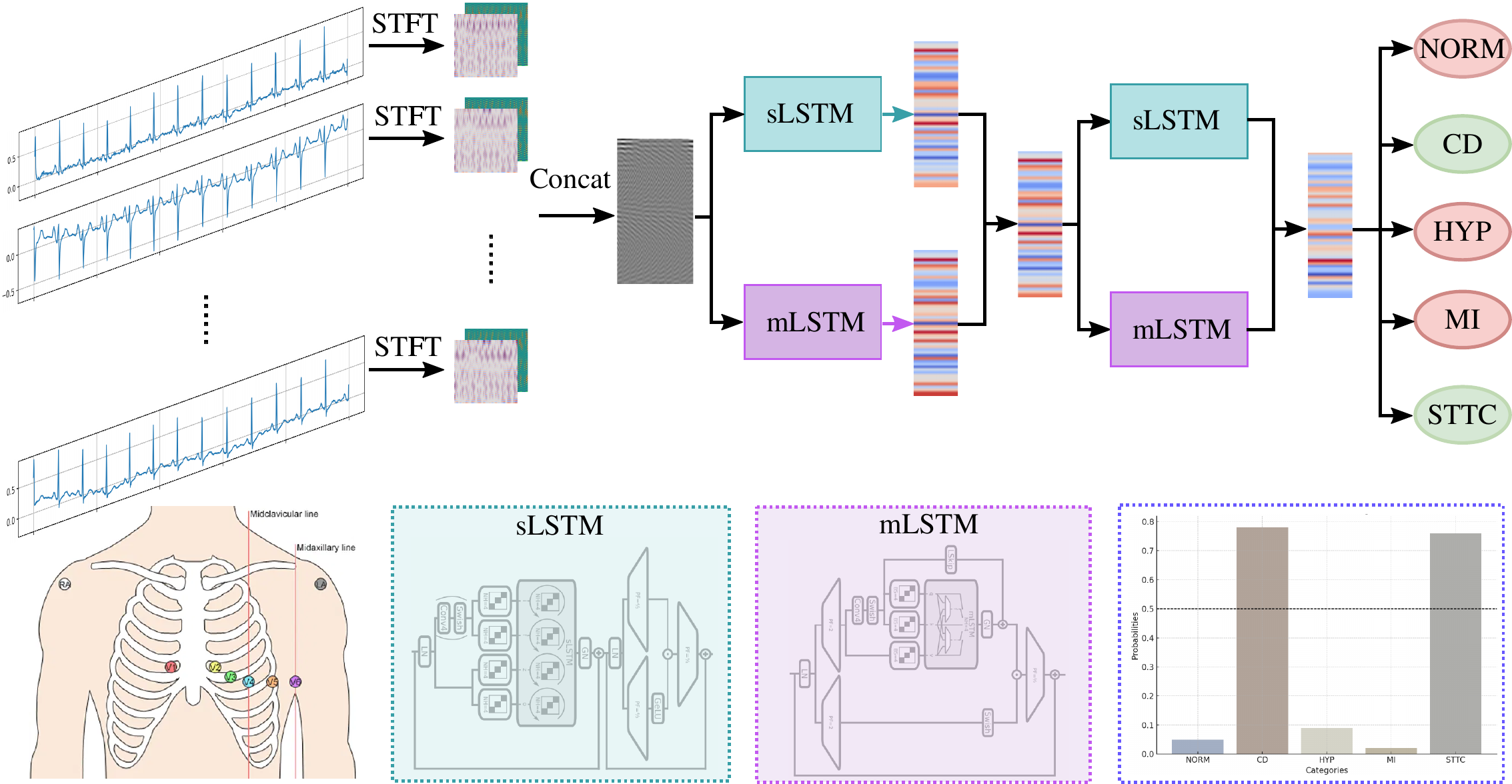}
  \caption{The architecture of the proposed xLSTM-ECG model begins with the input of 12-lead ECG waveforms $x_i$. After pre-processing, these waveforms are transformed into frequency domain features $\hat{x}_i$. These features are then fed into the layer fusion xLSTM blocks $\mathcal{M}$, which consist of the sLSTM (teal color) and mLSTM (lavender color) modules. The outputs from both modules $\hat{x}_{i}^{s}$ and $\hat{x}_{i}^{m}$ are merged to create a fusion feature $\hat{x}_{i}^{f}$. After passing through two blocks as shown in dashed rectangles, the final fusion feature undergoes average pooling to obtain a fixed-size vector $v_{i}$. This vector is then projected into probabilities for five different superclasses using five independent linear layers. Positive labels are indicated in \textcolor{green}{Green}, while negative labels are indicated in \textcolor{red}{Red}.}
  \label{fig:arch}
\end{figure*}

\subsection{Problem Formulation}

In this paper, we focus on 12-lead multiclass multi-label classification of 12-lead ECG signals. The ECG data $D = \{x_i, y_i\}^{N}_{i=1}$ where $x_i \in \mathbb{R}^{T \times 12}$ represents the 12-lead ECG data for the $i$-th sample, with $T$ being the number of time points; $y_i \in \{0, 1\}^C$ is the multiclass multi-label vector, where $C$ is the number of possible classes and each element indicates the presence $1$ or absence $0$ of a specific class for the $i$-th sample; $N$ is the total number of ECG data. The proposed method xLSTM-ECG is denoted as $\mathcal{F}$, which consists of two layer fusion xLSTM blocks $\mathcal{M}_0$ and $\mathcal{M}_1$ as shown in the dashed rectangle in Fig.~\ref{fig:arch}. Each block comprises sLSTM and mLSTM modules.

\subsection{Pre-processing}

ECG signals are generally long time-series data, capturing the heart's electrical activity over extended periods. Because of their length and complexity, directly feeding raw ECG data into a model can be computationally expensive and less efficient. Consequently, a preprocessing step is essential to enhance performance while reducing computational overhead.

ECG signals possess both stationary and non-stationary characteristics, underscoring the need for methods that effectively capture time-frequency information. Several signal processing approaches are available, including Wavelet Transform (WT), Empirical Mode Decomposition (EMD), and Short-Time Fourier Transform (STFT). Although WT is suitable for detecting transient features (like QRS complexes), it demands meticulous selection of the mother wavelet function and introduces additional hyperparameters~\cite{addison2005wavelet}. EMD adaptively decomposes signals into intrinsic mode functions~\cite{huang1998empirical}, but it is prone to noise sensitivity and lacks a standardized mathematical framework.

In comparison, STFT uses a fixed sliding window, allowing consistent, computationally efficient preprocessing across all leads. This uniformity is particularly beneficial for 12-lead ECG classification, where inter-lead consistency is crucial. Hence, we adopt STFT to transform each 12-lead ECG signal from the time domain into a time-frequency representation, thereby capturing both temporal and spectral patterns relevant to cardiac events.

Let $X \in \mathbb{R}^{T \times 12}$ represent a 12-lead ECG signal with $T$ time points. Each lead $X_l$ (for $l=1,2,\ldots,12$) contributes a unique spatial perspective of the heart’s electrical activity. Applying STFT yields:

\begin{equation}
X_{\text{STFT}} \in \mathbb{R}^{T' \times 12 \times F},
\end{equation}

where $T'$ is the number of time windows and $F$ is the number of frequency bins.

In our xLSTM-ECG model, we employ a dual-module design to capture both temporal dependencies and inter-lead interactions. For each time window $t$, the multi-lead features are represented as
$\mathbf{X}_t \in \mathbb{R}^{12 \times F}$. A modified LSTM processes these representations, producing a hidden state $h_t^{(l)}$ for each lead $l$. An adaptive fusion mechanism then integrates individual lead information:

\begin{equation}
h_t^{\text{fusion}} = \frac{1}{12} \sum_{l=1}^{12} h_t^{(l)} \;+\; \alpha \cdot 
\operatorname{Attention}\bigl(h_t^{(1)}, \dots, h_t^{(12)}\bigr),
\end{equation}

where \(\alpha\) is a learnable parameter that balances the contributions of the average hidden states and an attention-based aggregation of all leads. We also employ data augmentation (e.g., random masking) and a weighted loss function to mitigate lead-specific noise and address class imbalance.

The STFT’s time-frequency representation facilitates noise separation by isolating frequency bands, allowing filtering of high-frequency artifacts beyond the typical ECG bandwidth (commonly 0.5--40~Hz). It also aids in addressing baseline wander, usually a low-frequency artifact, through suitable high-pass filtering or detrending in the frequency domain. Applying STFT uniformly to all leads helps normalize data across channels, reducing the impact of lead-specific noise. Additionally, choosing appropriate window lengths and overlaps avoids smearing fast-changing events like the QRS complex, ensuring key morphological features (P, QRS, and T waves) remain distinguishable. Parameter tuning for the STFT is guided by domain knowledge and empirical testing on the PTB-XL dataset, striking a balance between preserving essential signal components and suppressing artifacts.

\subsection{xLSTM Modules}

The xLSTM architecture~\cite{beck2024xlstm} comprises two specialized extensions of the conventional LSTM: the Scalar Long Short-Term Memory (sLSTM) and the Matrix Long Short-Term Memory (mLSTM). Both modules address standard LSTM limitations by incorporating exponential gating and enhanced memory designs, enabling the learning and retention of information over longer time spans. Figure~\ref{fig:arch} provides an overview of these two modules: the sLSTM (teal color) targets fine-grained, short-term updates, whereas the mLSTM (lavender color) captures broader, long-term dependencies through a matrix-based memory mechanism. This dual-module framework is especially advantageous in 12-lead ECG classification, where rapid local variations (e.g., QRS complexes) coexist with slower, global shifts (e.g., ST-segment changes across multiple leads).

\textbf{sLSTM: Scalar Long Short-Term Memory.}
The sLSTM module introduces exponential gating functions and stabilization measures for more flexible, dynamic control over memory updates. Unlike a standard LSTM that primarily uses sigmoid and tanh activations, sLSTM applies an exponential function to regulate how new information enters the cell state and how old information is discarded. This design suits ECG signals, where abrupt waveform changes (e.g., arrhythmic beats) require swift adaptation.

In the sLSTM, each gate processes two inputs: the current time-step $x_t$ (potentially encompassing multi-lead features) and the previous hidden state $h_{t-1}$. The forget gate $f_t$ controls how much of the past cell state $c_{t-1}$ is retained, while the input gate $i_t$ decides how much new information is added. Together, these gates facilitate memory mixing, allowing the module to adjust stored information rapidly.

Denoting the sLSTM module by $\mathcal{M}_S$, its forward pass can be expressed as:
\begin{equation}
\hat{x}_{i}^{s} = \mathcal{M}_S (\hat{x}_i),
\end{equation}
where \(\hat{x}_i\) represents the input features at step \(i\), and \(\hat{x}_{i}^{s}\) is the short-term representation. Internally, the sLSTM updates its parameters as follows:

\textbf{Cell Status Update}:
\begin{equation}
c_t = f_t \odot c_{t-1} \;+\; i_t \odot z_t,
\end{equation}
where $c_t$ is the updated cell state, $f_t$ and $i_t$ are the forget and input gates, and $z_t$ is the candidate cell state. The symbol $\odot$ denotes elementwise multiplication.

\textbf{Standardizer Status Update}:
\begin{equation}
n_t = f_t \odot n_{t-1} + i_t,
\end{equation}
introduces a normalization variable $n_t$ that controls the scaling of the cell state over time, preventing exploding or vanishing values—an essential feature for processing ECG signals that vary in amplitude and frequency.

\textbf{Hidden State Update}:
\begin{equation}
h_t = o_t \odot \Bigl(\frac{c_t}{n_t}\Bigr),
\end{equation}
where $h_t$ is the hidden state, $o_t$ the output gate, and $n_t$ normalizes the cell state prior to output. This structure supports rapid adaptation to new information while maintaining hidden-state stability.

\textbf{Input and Forgetting}:
\begin{equation}
\begin{split}
  i_t &= \exp\bigl(W_i \, x_t + R_i \, h_{t-1} + b_i\bigr),\\
  f_t &= \exp\bigl(W_f \, x_t + R_f \, h_{t-1} + b_f\bigr),
\end{split}
\end{equation}
illustrates how the exponential gating mechanism selectively amplifies or diminishes inputs, enabling rapid responses to abrupt ECG changes. By blending exponential and sigmoid activations, the sLSTM gains finer control over state transitions, making it well-suited for short-lived ECG phenomena such as premature beats.

\textbf{mLSTM: Matrix Long Short-Term Memory.}
The mLSTM module expands LSTM memory capacity by using a matrix-based memory, affording a more expressive representation for capturing long-range dependencies. This is particularly relevant for 12-lead ECG data, where diagnostic indicators may emerge over extended intervals and multiple leads simultaneously.

In the mLSTM (Figure~\ref{fig:arch}, right), the cell state $c_t$ is updated via matrix operations, allowing the model to encode higher-order correlations. For instance, myocardial infarction (MI) might involve ST-segment deviations across several leads; the matrix-based update helps capture these cross-lead patterns over longer time spans.

Denoting the mLSTM module by $\mathcal{M}_M$, its forward pass is written as:
\begin{equation}
\hat{x}_{i}^{m} = \mathcal{M}_M (\hat{x}_i).
\end{equation}
Although \(\hat{x}_{i}^{m}\) also denotes updated features, the internal transitions differ from sLSTM:

\textbf{Cell Status Update}:
\begin{equation}
c_t = f_t \odot c_{t-1} \;+\; i_t \odot (v_t \, k_t^T),
\end{equation}
where \(v_t\) and \(k_t\) are matrix-like factors modeling covariance across different dimensions or leads. Such matrix operations enable richer data relationships, helping detect subtle changes spanning multiple ECG leads.

\textbf{Standardizer Status Update}:
\begin{equation}
n_t = f_t \odot n_{t-1} + i_t \odot k_t,
\end{equation}
serves a similar purpose to \(n_t\) in sLSTM but adapts it for matrix-based computations, preventing unbounded memory growth.

\textbf{Hidden State Update}:
\begin{equation}
h_t = o_t \odot \Bigl(c_t \, q_t \;\big/\; \max(q_t^T n_t,\, 1)\Bigr),
\end{equation}
yields the final hidden state by normalizing the matrix-updated cell state \(c_t\). Dividing by \(\max(q_t^T n_t, 1)\) ensures numerical stability, which is crucial for high-dimensional data in 12-lead ECG processing.

By utilizing a matrix memory structure, the mLSTM effectively models complex inter-lead dependencies—useful for detecting global phenomena like multi-lead ST elevations or conduction delays spanning longer intervals. This complements the sLSTM’s emphasis on short-term details, forming a comprehensive ECG feature-extraction pipeline. 

Figure~\ref{fig:arch} highlights the primary distinctions between sLSTM (left) and mLSTM (right). In practice, both modules receive the same input sequence, and their outputs are later fused. This fusion combines the sLSTM’s focus on local, short-term phenomena with the mLSTM’s broader, cross-lead context, significantly enhancing the traditional LSTM’s capacity to manage the rich temporal and spatial complexity found in 12-lead ECG signals. The next section provides a detailed overview of the feature-fusion approach.

\subsection{Feature Fusion Strategies}

\begin{figure}[t!]
  \centering
  \includegraphics[width=0.5\linewidth]{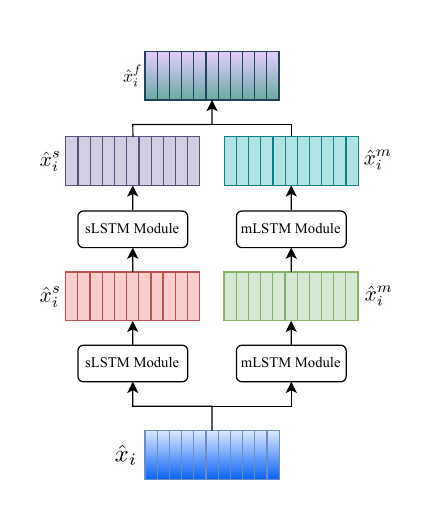}
  \caption{Sequential fusion strategy: features are integrated at the final layer of both sLSTM and mLSTM modules.}
  \label{fig:fusionA}
\end{figure}

\begin{figure}[t!]
  \centering
  \includegraphics[width=0.5\linewidth]{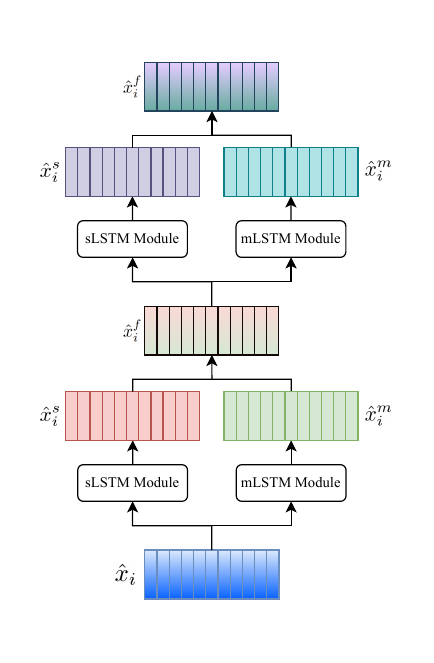}
  \caption{Layer fusion strategy: features are integrated during each layer of both sLSTM and mLSTM modules.}
  \label{fig:fusionB}
\end{figure}

Feature fusion plays a pivotal role in enhancing the xLSTM architecture by integrating information across various layers and modules. In multi-lead ECG classification, combining features at different stages helps capture both short-term, local dynamics and long-term inter-lead dependencies. We explore two fusion strategies: the \emph{sequential fusion strategy} and the \emph{layer fusion strategy}, illustrated in Figures~\ref{fig:fusionA} and~\ref{fig:fusionB}.

\textbf{Sequential Fusion Strategy.}
In the \emph{sequential fusion strategy} (Figure~\ref{fig:fusionA}), features are combined only at the final layer of the sLSTM and mLSTM modules. Each module processes the input \(\hat{x}_i\) independently, producing an sLSTM feature \(\hat{x}_{i}^{s}\) and an mLSTM feature \(\hat{x}_{i}^{m}\). At the last layer, these outputs are merged as follows:
\begin{equation}
\hat{x}_{i}^{f} = \frac{\hat{x}_{i}^{s} + \hat{x}_{i}^{m}}{2}.
\end{equation}
This approach leverages each module’s specialized strengths: the sLSTM excels at short-term ECG variations (e.g., QRS complexes), while the mLSTM focuses on longer-term patterns or inter-lead correlations. By delaying fusion until the end, the sequential strategy simplifies implementation and ensures minimal intermediate interaction between modules.

\textbf{Layer Fusion Strategy.}
In contrast, the \emph{layer fusion strategy} (Figure~\ref{fig:fusionB}) merges features at every layer of both sLSTM and mLSTM. After each layer \(l\), the outputs are fused (e.g., via averaging) before proceeding to the next layer:
\begin{equation}
\hat{x}_{i,l}^{f} = \frac{\hat{x}_{i,l}^{s} + \hat{x}_{i,l}^{m}}{2}.
\end{equation}
Here, \(\hat{x}_{i,l}^{s}\) and \(\hat{x}_{i,l}^{m}\) are the layer-\(l\) outputs of the sLSTM and mLSTM, respectively. This continuous exchange of features enables the model to refine both local (sLSTM-driven) and global (mLSTM-driven) representations at each layer.

By repeatedly blending sLSTM and mLSTM outputs, the layer fusion strategy offers the following benefits:
\begin{itemize}
    \item \textbf{Iterative Feature Refinement:} Early-stage features from each module are continuously improved by leveraging the complementary representations of the other module.
    \item \textbf{Enhanced Inter-Lead Interactions:} Frequent fusion promotes stronger modeling of global dependencies (mLSTM) alongside local details (sLSTM).
    \item \textbf{Robustness to Noise:} Repeatedly integrating information from both modules helps mitigate the impact of lead-specific or transient noise.
\end{itemize}

Although more sophisticated fusion methods (e.g., attention mechanisms) might further optimize results, we use simple averaging for its computational efficiency and stable performance, as demonstrated by preliminary experiments on the PTB-XL dataset. Simple averaging imposes minimal additional parameters, reducing the risk of overfitting and ensuring a balanced representation of both short-term and long-term features.

In summary, the sequential fusion strategy preserves each module’s specialized pipeline until the final stage, while the layer fusion strategy integrates features more frequently to create an intertwined flow of information. Empirical evidence suggests that layer fusion often leads to richer representations of multi-lead ECG data, making it particularly effective for capturing the diverse morphological variations seen in real-world clinical settings.

\subsection{Multi-label Prediction}
To obtain a fixed-dimensional representation, we employ average pooling on the fused ECG data $\hat{x}_{i}^{f}$ to aggregate into a compact and fixed-size vector $v_i$ by averaging the features across the sequence. Mathematically, this operation for the $i$-th sample can be expressed as:

\begin{equation}
    v_{i} = AveragePool(\hat{x}_{i}^{f})
\end{equation}

Following the pooling operation, we project this vector onto a space corresponding to $C$ categories. This is simply achieved through $C$ linear layers. The multi-label predictions are obtained as follows:

\begin{equation}
    p_{i}^{j} = W_j v_{i} + b_j
\end{equation}

where $p_{i}^{j}$ is the predicted probability on $i$-th sample at $j$-th category, $W_j$ and $b_j$ represent the weight and bias of the linear layer corresponding to the $j$-th category, respectively.

\subsection{Loss}

Since we have obtained the predicted probabilities $p_{i}^{j}$ where $j \in \{1, 2, ..., C\}$. We utilize Binary Cross-Entropy Loss (BCELoss) to compute the loss as follows:

\begin{equation}
  L_{BCE} = -\frac{1}{N} \sum_{i}^{N} \sum_{j}^{C} \left( y_{ij} \log p_{i}^{j} + (1 - y_{ij}) \log (1 -p_{i}^{j}) \right)
\end{equation}

where $y_{ij}$ is the groundtruth of $i$-th sample at $j$-th category, which is binary either 1 (positive) or 0 (negative).

\subsection{Data Augmentation}
Random masking is utilized as a data augmentation technique to enhance the diversity and robustness of our ECG signal training data. This technique involves selecting random segments within the ECG signals and masking the data in these areas. By introducing these controlled occlusions, the model is encouraged to learn more generalized features because it cannot depend solely on specific segments of the input signals for predictions. This approach helps simulate real-world scenarios where parts of the ECG signals may be corrupted or missing, thereby improving the model's ability to handle incomplete or noisy data. Additionally, random masking acts as a form of regularization, reducing the risk of overfitting by preventing the model from relying too heavily on specific patterns in the training data. Consequently, this data augmentation strategy is crucial in enhancing the model's robustness and generalizability for ECG signal analysis.

\section{Experiments}

\subsection{Dataset}

We utilize the PTB-XL dataset, which is the largest publicly accessible clinical ECG dataset to date, as described by~\cite{wagner2020ptb}. It comprises 21,837 clinical 12-lead ECG records, each 10 seconds in length, collected from 18,885 patients. Of these patients, 52\% were male and 48\% were female. The dataset provides two types of sampling frequencies: 100Hz and 500Hz. In this paper, we focus on the 100Hz data. The ECG records are categorized into three non-reciprocal categories: diagnosis, form, and rhythm. For the diagnostic labels, there are 5 superclasses and 24 subclasses. Our study focuses on the multiclass multi-label classification task involving 5 diagnostic superclasses: \textbf{NORM} (Normal ECG), \textbf{CD} (Conduction Disturbance), \textbf{HYP} (Hypertrophy), \textbf{MI} (Myocardial Infarction), and \textbf{STTC} (ST/T Change). The statistics for the five superclasses in the PTB-XL dataset are presented in Tab.~\ref{tab:data}. Additionally, we utilize the recommended train-set splits from the dataset in~\cite{wagner2020ptb}, which assigns records to one of ten cross-validation folds. The tenth fold is used as the test set, the ninth fold as the validation set, and the first eight folds as the training set.

\begin{table}[h!]
    \caption{Statistics of the PTB-XL dataset on five superclasses.}
    \centering
    \small
    \begin{tabular}{ccc}
    \toprule
    \textbf{Class} & \textbf{Description} & \textbf{Number of Records}\\
    \midrule
    NORM & Normal ECG & 7,185\\
    CD & Conduction Disturbance & 3,232\\
    HYP & Hypertrophy & 815\\ 
    MI & Myocardial Infarction & 2,936\\
    STTC & ST/T Change & 3,064\\
    \bottomrule
    \end{tabular}
    \label{tab:data}
\end{table}

\subsection{Metrics}

In this paper, we employ two widely recognized performance metrics to assess the effectiveness of our method: Accuracy and macro averaging Area Under the RoC Curve (AUC). The definition of Accuracy is given by the following equation:
\begin{equation}
    Accuracy = \frac{TP + TN}{TP + FP + TN + FN}
\end{equation}
where TP (True Positives), FP (False Positives), TN (True Negatives), and FN (False Negatives) represent the counts of each respective outcome in the classification tasks. This metric quantifies the overall proportion of correct predictions (both true positives and true negatives) out of the total number of cases examined. Both Accuracy and AUC are metrics where higher values indicate better performance.

\subsection{Implementation Details}

In our experimental framework, we use Python 3.11 and PyTorch 2.3, with computational acceleration provided by CUDA 12.2 on an NVIDIA A40 GPU. We process data in batches of 1024. Training commences with an initial learning rate of 0.0002, which is systematically reduced by 20\% every two epochs by a scheduler. To mitigate overfitting, training automatically halts after five consecutive epochs without performance gains using an early stopping protocol. We perform feature extraction using the Short-Time Fourier Transform (STFT), with a window size of 64 and a hop length of 16, effectively delineating the time-frequency characteristics of ECG signals. Our model training employs the Adam optimizer and uses binary cross-entropy for loss calculation, which is apt for tasks involving multiclass and multi-label classification. Additionally, we apply a dropout rate of 0.5 and enhance model robustness through the use of random masking as a data augmentation strategy. Detailed strategies for extensive hyperparameter optimization are elaborated in the Hyperparameter Search section~\ref{sec:hyper}.

\subsection{Hyperparameter Search}
\label{sec:hyper}

During the hyperparameter search experiments, each test is consistently evaluated using a validation set with a constant learning rate of 0.0002 and an early stopping protocol activated after two epochs without progress. This method ensures a precise and unbiased assessment of each hyperparameter's impact.

In our first experiment, we focus on identifying the optimal number of Fast Fourier Transform (FFT) features for the Short-Time Fourier Transform (STFT). This is crucial for accurately capturing the time-frequency dynamics of ECG signals. We explore four configurations: 240, 360, 480, and 512 FFT features. This experiment allows us to ascertain the number of features that offer the best clarity without incurring excessive noise or computational demands. As reported in Table~\ref{tab:nfft}, the configuration with 480 FFT features emerged as the most effective that achieves the highest accuracy and AUC on the validation set. This indicates that 480 FFT features strike an optimal balance between detailed information capture and performance efficiency, making it our preferred choice.

\begin{table}[h!]
    \caption{Hyperparameter search on number of Fast Fourier Transform.}
    \label{tab:nfft}
    \centering
    \small
    \begin{tabular}{ccc}
    \toprule
    \textbf{N\_FFT} & \textbf{Accuracy (\%)}  &  \textbf{AUC (\%)}\\
    \midrule
    240 & 87.29 & 90.66\\
    360 & 87.52 & 90.69\\
    \textbf{480} & \textbf{87.72} & \textbf{91.12}\\
    512 & 87.38 & 90.79 \\
    \bottomrule
    \end{tabular}
\end{table}

\begin{table}[h]
    \caption{Hyperparameter search on window mask ratio, which determines the proportion of the ECG signals that are masked.}
    \label{tab:winmask}
    \centering
    \small
    \begin{tabular}{ccc}
    \toprule
    \textbf{Window Mask Ratio} & \textbf{Accuracy (\%)}  &  \textbf{AUC (\%)}\\
    \midrule
    0.1 & 87.11 & \underline{90.96}\\
    0.2 & \underline{87.19} & \textbf{91.00}\\
    0.3 & 87.10 & 90.95\\
    0.4 & 87.10 & 90.95\\
    0.5 & \textbf{87.25} & 90.87\\
    \bottomrule
    \end{tabular}
\end{table}

Next, we seek to identify the optimal masking ratio for our random masking data augmentation technique by evaluating various ratios, specifically 0.1, 0.2, 0.3, 0.4, and 0.5. This technique involves randomly concealing sections of the input data, enabling the model to learn to make predictions based on incomplete or obscured information, thereby enhancing its adaptability to unseen task variations. According to the results presented in Table~\ref{tab:winmask}, a masking ratio of 0.2 achieves a favorable balance, yielding the second-highest accuracy and the highest AUC. This ratio effectively encourages the model to develop robust features by concealing sufficient data to challenge its predictive capabilities, yet it retains enough information to prevent significant loss of detail. Thus, we select a masking ratio of 0.2 for its optimal balance of data utility and augmentation impact.

Finally, we investigate the optimal frequency for applying our random masking data augmentation technique during model training. Using too high a probability might hinder the model from learning useful features from the input data, whereas too low a probability could lead the model to overfit the training set and thus introduce a substantial bias between training and test performances. We evaluate the probabilities ranging from 0.6 to 1.0 to determine the ideal frequency of augmentation application to data batches. This strategic choice affects both the diversity of the training examples and the stability of the learning process. As reported in Table~\ref{tab:augment}, a probability of 0.8 is found to be optimal, achieving the second-highest accuracy and AUC. This frequency, augmenting 80\% of the data batches, effectively enhances model resilience and performance on unseen data, striking a crucial balance between robustness and training consistency.

\begin{table}[h]
    \caption{Hyperparameter search on augmentation probability.}
    \label{tab:augment}
    \centering
    \small
    \begin{tabular}{ccc}
    \toprule
    \textbf{Probability} & \textbf{Accuracy (\%)}  &  \textbf{AUC (\%)}\\
    \midrule
    0.6 & 87.48 & 91.12\\
    0.7 & 87.39 & \textbf{91.23}\\
    0.8 & \underline{87.62} & \underline{91.17}\\
    0.9 & \textbf{87.72} & 91.12\\
    1.0 & 87.53 & 91.09\\
    \bottomrule
    \end{tabular}
\end{table}

\subsection{Ablation Study}


In this study, we investigate the effect of the sLSTM and mLSTM modules through ablation experiments, focusing on their impact on model performance, specifically accuracy and AUC. The Tab.~\ref{tab:ablation} summarize the performance outcomes for various module configurations and fusion strategies. Initially, we assess the individual contributions of the sLSTM and mLSTM modules by comparing their standalone performance with the results achieved by combining both modules using different fusion approaches.

The experimental results show that using the sLSTM and mLSTM modules alone achieves 87.29\% accuracy and 90.79\% AUC, and 87.53\% accuracy and 90.96\% AUC, respectively. These findings suggest that both modules exhibit similar performance, with mLSTM slightly outperforming sLSTM. When the sLSTM and mLSTM modules are combined using the sequential fusion method, performance improves marginally, reaching 87.64\% accuracy and 90.97\% AUC. This result demonstrates that sequentially fusing the two modules provides a modest enhancement in model performance. However, the most significant improvement is observed with the layer fusion approach, where the accuracy and AUC reach their highest values at 87.78\% and 91.10\%, respectively. This outcome underscores the complementary nature of the sLSTM and mLSTM modules when combined via the layer fusion method, resulting in further enhancement of the model's performance.

In conclusion, the ablation experiments provide clear evidence that the combination of the sLSTM and mLSTM modules, along with the appropriate fusion strategy, significantly enhances the model's accuracy and AUC. Notably, the layer fusion approach yields the best results, demonstrating its effectiveness in improving performance for the ECG classification task.

\begin{table}[h]
    \caption{Ablation study on sLSTM and mLSTM modules. }
    \label{tab:ablation}
    \centering
    \small
    \scalebox{0.95}{
    \begin{tabular}{cccccc}
    \toprule
    \textbf{sLSTM} & \textbf{mLSTM} & \textbf{Fusion Type} & \textbf{Accuracy (\%)}  &  \textbf{AUC (\%)}\\
    \midrule
    \checkmark & $-$ & $-$ & 87.29 & 90.79 \\
    $-$ & \checkmark  & $-$ & 87.53 & 90.96 \\
    \checkmark & \checkmark & Sequential & 87.64 & 90.97 \\
    \checkmark & \checkmark & \textbf{Layer} & \textbf{87.78} & \textbf{91.10}\\
    \bottomrule
    \end{tabular}
    }
\end{table}

\subsection{Main Results and Analysis}


\subsubsection{Confusion Matrices}

\begin{figure}[h!]
  \centering
  \includegraphics[width=0.98\linewidth]{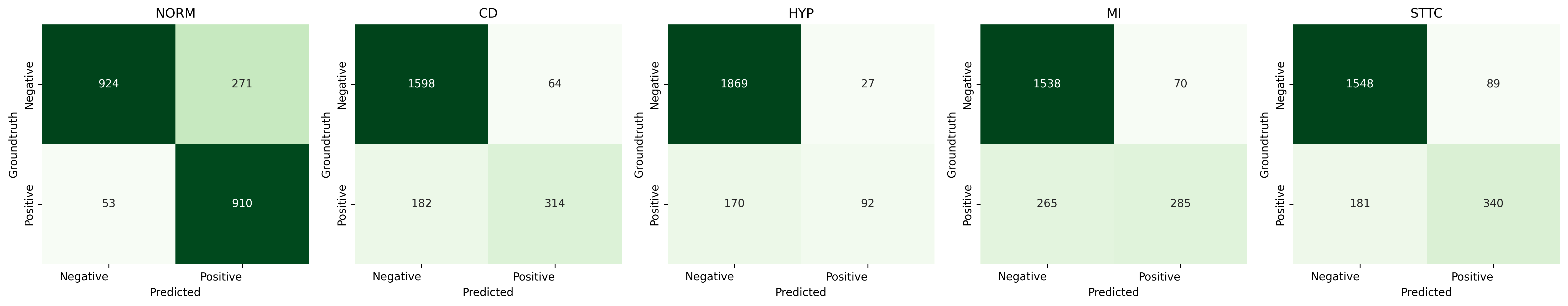}
  \caption{Confusion matrices corresponding to the five categories: NORM, CD, HYP, MI, and STTC.}
  \label{fig:confusion}
\end{figure}

Since this is a multi-class multi-label classification task, Fig.~\ref{fig:confusion} presents the confusion matrices for the NORM, CD, HYP, MI, and STTC categories. The model demonstrates strong performance in identifying normal ECGs (NORM), with numerous true positives and true negatives and relatively few misclassifications. It also performs reasonably well on conduction disturbances (CD) and ST/T changes (STTC), exhibiting moderate sensitivity and good specificity. In contrast, the detection of hypertrophy (HYP) and myocardial infarction (MI) remains more challenging, as indicated by a higher incidence of false negatives and fewer true positives. The results illustrate the strengths and limitations of the proposed model in this multi-class, multi-label classification task. We study the findings for each category as follows.

For Normal ECGs (NORM), the model achieves strong performance in identifying normal ECG patterns, with a high number of true positives (910) and true negatives (924). Misclassifications are relatively rare, with 53 false negatives and 271 false positives. These results indicate excellent sensitivity and specificity, reflecting the model's robust ability to distinguish normal cases from abnormal ones.

For Conduction Disturbance (CD), the model shows balanced performance, with 314 true positives and 1598 true negatives, along with 182 false negatives and 64 false positives. The high specificity suggests that the model effectively minimizes false positives, but the moderate false negative rate indicates some difficulty in detecting true conduction disturbances, likely due to the heterogeneous ECG patterns associated with CD.

Hypertrophy (HYP) detection remains a significant challenge for the model. Although it achieves 1869 true negatives, it identifies only 92 true positives, with a substantial number of false negatives (170). This results in low sensitivity, which highlights the model's difficulty in recognizing the subtle ECG features indicative of hypertrophy. False positives are minimal, demonstrating robust specificity.

Similarly, the Myocardial Infarction (MI) classification shows limitations, with 285 true positives and 1538 true negatives, but a high number of false negatives (265). This indicates that subtle deviations from the ECG associated with myocardial infarction are difficult to capture. Despite robust specificity, the low sensitivity underscores the need for improved feature extraction to reduce missed positive cases.

The model performs moderately well in identifying ST/T changes (STTC), with 340 true positives and 1548 true negatives. However, 181 false negatives and 89 false positives suggest room for improvement in sensitivity. The results indicate that, while the model reliably identifies most cases, it occasionally misclassifies borderline instances.

The model excels in detecting normal ECGs (NORM), achieving high sensitivity and specificity. It also demonstrates robust specificity in CD and STTC detection, effectively minimizing false positives. Detecting HYP and MI remains problematic, as evidenced by higher false-negative rates and fewer true positives. These limitations point to difficulties in capturing subtle or less distinct ECG features, which may require advanced feature extraction techniques or further model optimization. Overall, the confusion matrices provide valuable insight into the model's diagnostic capabilities, emphasizing its effectiveness in certain categories, while highlighting areas for refinement.

\subsubsection{Label Co-occurence Matrices}



\begin{figure}[h!]
  \centering
  \begin{subfigure}[b]{0.48\textwidth}
  \centering
  \includegraphics[width=\textwidth]{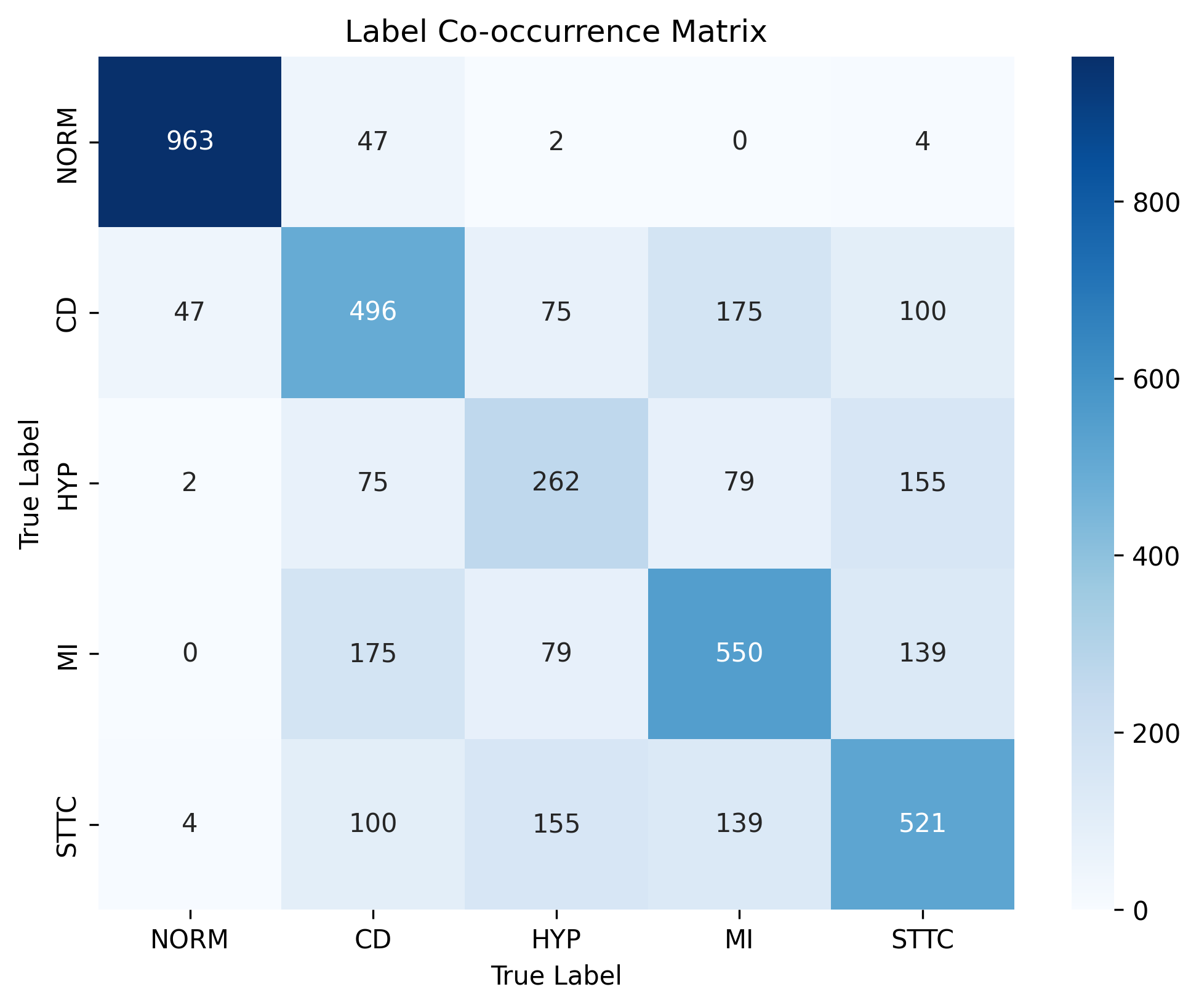}
  \caption{True vs. true label co-occurrence matrix.}
  \end{subfigure}
\hfill
  \begin{subfigure}[b]{0.48\textwidth}
  \centering
  \includegraphics[width=\textwidth]{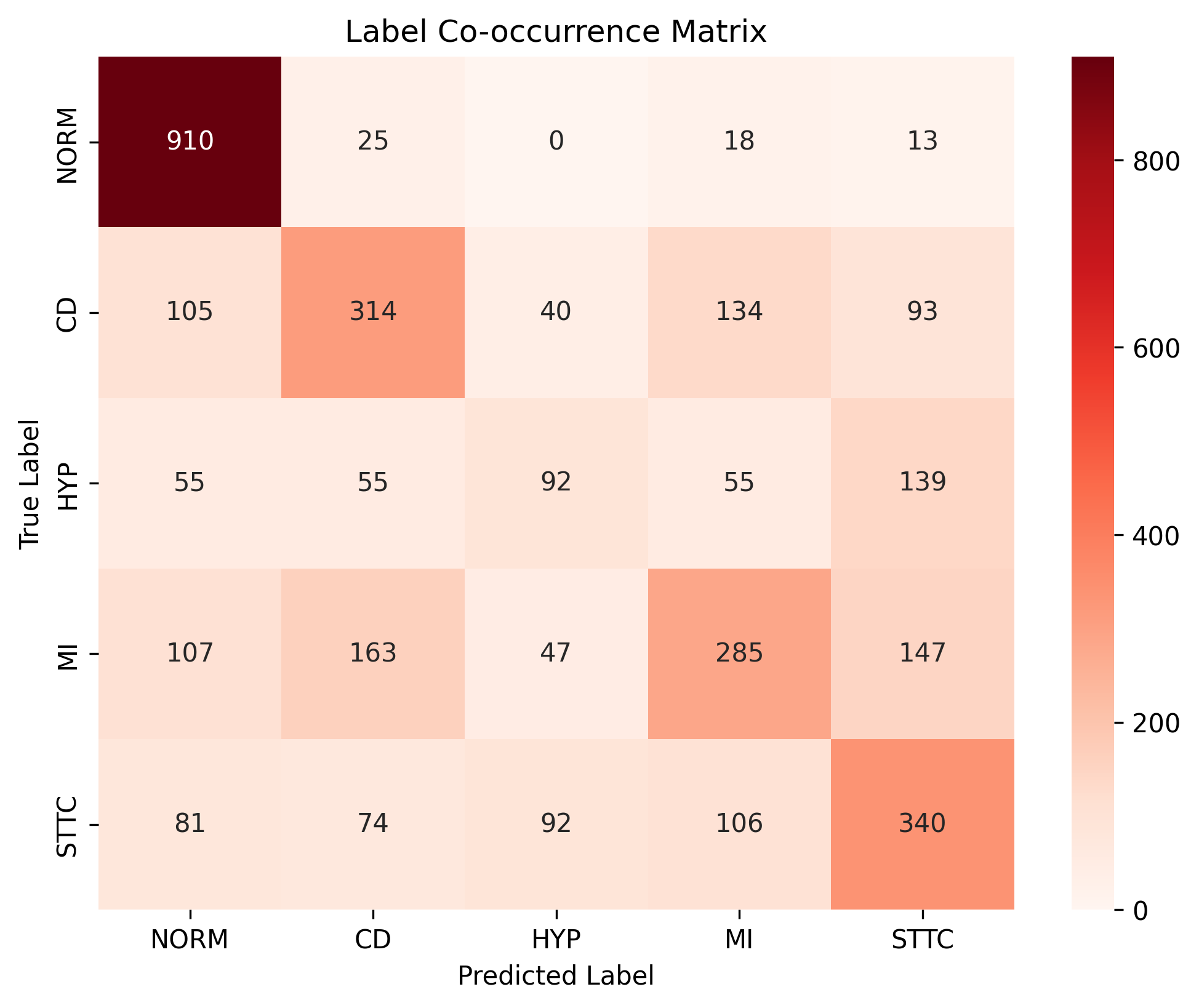}
  \caption{True vs. predicted label co-occurrence matrix.}
  \end{subfigure}
  \caption{Label co-occurrence matrices for multi-class, multi-label classification: (a) shows the ground-truth co-occurrence matrix, and (b) shows the predicted co-occurrence matrix.}
  \label{fig:cooccurence}
\end{figure}

We examine the Label Co-occurrence Matrix (LCM) for the multi-class multi-label classification task as illustrated in Fig.~\ref{fig:cooccurence}. In this figure, the rows and columns represent true labels in matrix (a), and true vs. predicted labels in matrix (b), with each cell indicating how often two categories co-occur. In the ground-truth co-occurrence matrix (blue matrix), NORM predominantly appears alone (963 instances), but also co-occurs with CD (47), HYP (2), and STTC (4). While multi-label classification permits the coexistence of multiple conditions, these unexpected pairings of NORM with abnormal categories suggest label noise or inconsistencies in the annotation process.

In the ground-truth co-occurrence matrix (blue matrix), the CD label has 496 self-occurrences, reflecting a significant number of samples explicitly labeled with conduction disturbances. Additionally, the CD label frequently co-occurs with HYP (75), MI (175), and STTC (100), suggesting that conduction disturbances often coexist with hypertrophy, myocardial infarction and ST/T changes, respectively. The HYP label has 262 self-occurrences, indicating that a substantial number of samples are explicitly labeled with hypertrophy. Notably, HYP frequently co-occurs with CD (75), MI (79), and STTC (155), highlighting a potential association between hypertrophy and other cardiac conditions. The MI label shows a self-occurrence count of 550, indicating that many samples are explicitly labeled with myocardial infarction. It also exhibits significant co-occurrence with CD (175) and STTC (139), suggesting that myocardial infarction is often associated with conduction disturbances and ST/T wave changes. The STTC label appears 521 times as a self-occurrence and frequently co-occurs with CD (100), HYP (155), and MI (139). This indicates a strong relationship between ST/T wave changes and other cardiac abnormalities, particularly myocardial infarction and hypertrophy. The LCM highlights complex relationships between different cardiac conditions and suggests the potential presence of labeling noise or inconsistencies. For example, the co-occurrences of NORM with abnormal labels could result from errors in the labeling process. Similarly, the frequent co-occurrences of labels such as CD, HYP, MI, and STTC underscore the overlapping nature of cardiac conditions, which may complicate classification tasks.

These findings are critical for improving the data labeling process by identifying areas of inconsistency. Addressing such issues can enhance the quality of the dataset and, in turn, improve the performance of multi-class, multi-label classification models. Furthermore, the observed co-occurrence patterns may provide valuable insights into the relationships between different cardiac conditions, which could inform future model design and diagnostic strategies.

Examining the true vs. predicted co-occurrence matrix (the red matrix in Fig.~\ref{fig:cooccurence}) reveals a pattern similar to the ground-truth distribution: the model tends to assign the highest probability to each label’s primary class, reflecting the co-occurrence tendencies seen in the data. For CD and STTC, predictions generally align with their observed multi-label distributions. However, HYP remains problematic, as many HYP-positive samples are misclassified as STTC, and some MI instances are identified as NORM, indicating that these distinctions are harder to capture. Fig.~\ref{fig:cooccurence}(b) provides a detailed Label Co-occurrence Matrix (LCM) for the true vs. predicted scenario, offering insights into the frequency and relationships among label co-occurrences and highlighting these key observations.

The NORM label shows the highest self-occurrence, with 910 instances, indicating that most normal samples are correctly identified without co-labeling with abnormal conditions. However, there are some co-occurrences with other labels, such as CD (25), MI (18), and STTC (13), while none were observed with HYP. These co-occurrences could point to potential labeling inconsistencies or noise within the dataset. The CD label has 314 self-occurrences, representing a substantial number of samples explicitly labeled with conduction disturbances.
Notable co-occurrences include 134 instances with MI and 93 instances with STTC, suggesting a frequent overlap between conduction disturbances, myocardial infarction, and ST/T wave changes. HYP exhibits 92 self-occurrences, indicating a smaller subset of samples explicitly labeled with hypertrophy. It frequently co-occurs with CD (40) and STTC (139), highlighting potential associations between hypertrophy and other cardiac abnormalities. The MI label has a self-occurrence count of 285, reflecting a significant number of samples labeled with myocardial infarction.
It shows strong co-occurrence with CD (134) and STTC (147), underlining a frequent association between myocardial infarction, conduction disturbances, and ST/T wave changes. The STTC label appears 340 times as a self-occurrence, reflecting its frequent identification as an independent condition.
It co-occurs prominently with MI (147) and HYP (13), suggesting a strong interrelationship between ST/T wave changes, myocardial infarction, and hypertrophy.

To further understand the challenges of multi-label ECG classification, we conducted a detailed analysis of interclass overlapping. Our results reveal that a notable proportion of samples exhibit overlapping labels. For instance, a subset of recordings labeled as NORM also carries abnormal annotations such as CD or STTC. We quantified the overlapping by computing co-occurrence rates between each pair of classes. The analysis shows that classes with high overlap, particularly those involving NORM and abnormal categories, correlate with a reduction in sensitivity for minority classes such as HYP and MI. This interclass overlapping likely reflects inherent ambiguities in the ECG signal characteristics or inconsistencies in the labeling process. Recognizing these overlaps is critical, as they may contribute to misclassification errors and highlight the need for future work to incorporate advanced feature extraction or loss function adjustments that explicitly account for label correlations. The observed interclass overlapping, as evidenced by our co-occurrence analysis, suggests that some misclassifications may stem from ambiguous or overlapping features among classes. Addressing this challenge could involve exploring adaptive loss functions or enhanced fusion strategies, which will be a focus of future research.

The true vs. predicted LCM reveals intricate interdependencies and frequent co-occurrences among various cardiac conditions. In particular, the common co-labeling of CD, MI, HYP, and STTC emphasizes the overlapping nature of these abnormalities, making it harder for the classification model to accurately distinguish one condition from another. This complexity is especially evident when detecting HYP and MI, and will guide our future work.


\subsubsection{Comparison with SoTA}

Finally, in Tab.~\ref{tab:results}, we compare the proposed method against state-of-the-art approaches. The results indicate that the proposed xLSTM-based method outperforms competing solutions in both accuracy and AUC. In particular, it delivers a substantial performance gain over existing LSTM-based models, achieving a 45\% improvement in accuracy and a slight AUC increase compared to HLSTM~\cite{jyotishi2020attention}, and surpassing LSTM~\cite{martin2021real} by 4\% in accuracy and 1\% in AUC.

Our model attains first place in both Accuracy and AUC, achieving 87.59\% and 91.33\%, respectively. ECGNet achieves the second highest Accuracy at 87.35\%, while Resnet101 attains the third highest Accuracy at 86.78\%, both accompanied by relatively high AUC values of 91.01\% and 89.52\%, respectively.

ASTLNet achieves the second best AUC at 91.31\%, but only attains an Accuracy of 62.69\%. Similarly, HLSTM achieves the third best AUC at 91.26\%, but its Accuracy is as low as 60.61\%. While their AUC values are comparable to our proposed method’s AUC of 91.33\%, their Accuracy scores are significantly lower than our result of 87.59\%. Thus, both ASTLNet and HLSTM, despite exhibiting a high AUC but low Accuracy, tend to assign higher scores to the minority class compared to the majority class, which could yield a high AUC in an imbalanced category task.

\begin{table}[h]
    \caption{Comparison with the state of the art on PTB-XL dataset.}
    \label{tab:results}
    \centering
    \small
    \begin{tabular}{cccc}
    \toprule
    \textbf{Method} & \textbf{Year} & \textbf{Accuracy (\%)}  &  \textbf{AUC (\%)}\\
    \midrule
    VGG16~\cite{simonyan2014very} & 2014 & 62.14 & 90.93\\
    Resnet50~\cite{he2016deep} & 2016 & 60.75  & 91.01\\
    ECGNet~\cite{murugesan2018ecgnet} & 2018 & \underline{87.35}  & 91.01 \\
    Seq2Seq~\cite{mousavi2019inter} & 2019 & 84.19  & 86.54 \\
    ATICNN~\cite{yao2020multi} & 2020 & 62.32 & 91.05 \\
    DMSFNet~\cite{wang2020deep} & 2020 & 62.51 &   90.72 \\
    HLSTM~\cite{jyotishi2020attention} & 2020 & 60.61 &  91.26 \\
    ASTLNet~\cite{jyotishi2023attentive} & 2020 & 62.69 &  \underline{91.31} \\
    Resnet101~\cite{reddy2021imle} & 2021 & 86.78 &  89.52 \\
    LSTM~\cite{martin2021real} & 2021 & 84.17  & 90.3 \\
    METS~\cite{li2024frozen} & 2023 & 84.2 & $-$\\ 
    MVKT-ECG~\cite{qin2023mvkt} & 2023 & 82.2 & 84.3 \\
    ETP~\cite{liu2024etp} & 2024 & $-$ & 83.5 \\
    \midrule
    \textbf{Proposed} & \textbf{2025} & \textbf{87.59}  & \textbf{91.33}\\
    \bottomrule
    \end{tabular}
\end{table}

Most methods in Table~\ref{tab:results} achieve balanced performance on Accuracy and AUC; however, our xLSTM-based model demonstrates superior results on both metrics. This dual strength indicates not only a high degree of correct classification across a broad set of ECG samples (high Accuracy) but also a strong ability to differentiate between classes at multiple decision thresholds (high AUC). Such robust performance under imbalanced scenarios highlights the model’s enhanced sensitivity, specificity, and overall clinical utility.

The improved performance of the proposed xLSTM-ECG architecture can be largely attributed to its extended memory features. While standard LSTMs often encounter vanishing gradients and rely on limited memory designs, the xLSTM framework incorporates both a scalar LSTM (sLSTM) and a matrix LSTM (mLSTM). The sLSTM module excels at short-term, local temporal changes such as QRS complex transitions, whereas the mLSTM module employs a matrix-based memory update to maintain long-term dependencies and model cross-lead relationships—vital for interpreting 12-lead ECG patterns. This setup yields richer, more comprehensive feature representations compared to traditional LSTM or Transformer-based networks, which, despite capturing global dependencies, may demand considerable computing power and can be less adaptable to nuanced inter-lead variations.

Moreover, our approach effectively tackles obstacles intrinsic to ECG data, including imbalanced class distributions wherein minority classes (e.g., Hypertrophy and Myocardial Infarction) are underrepresented. By harnessing an extended memory component, the xLSTM-ECG model better identifies subtle pathological features, thus reducing the rate of misclassification. In addition, the gating strategies in both sLSTM and mLSTM help mitigate common noise artifacts such as baseline wander, further enhancing the signal quality for classification. Coupled with a robust fusion mechanism, this design considerably improves accuracy while achieving AUC values on par with other high-performing methods.

Although macro-average AUCs are generally close across the evaluated models, the higher overall Accuracy delivered by our xLSTM-ECG model underscores its effectiveness in accurately classifying a greater portion of ECG traces, particularly from minority classes. By offering an extended memory capacity, noise resilience, and efficient inter-lead integration, our method proves more resilient to the inherent variability and class imbalance in PTB-XL. From a clinical standpoint, increased Accuracy implies that essential diagnostic indicators are both preserved and leveraged effectively, providing a tangible advantage in screening environments where classification precision is paramount.

By introducing architectural improvements and advanced training protocols, this work sets a new performance benchmark in automated ECG classification. Enhancing the correct detection of normal and abnormal ECG patterns effectively reduces misdiagnoses and underdiagnoses, promoting timely evidence-based clinical decisions. Consequently, our approach presents significant potential for advancing patient care in the context of life-threatening cardiac conditions.

In summary, the xLSTM-ECG model extends standard LSTM capabilities by integrating a dual-module configuration with feature fusion. Extensive experimentation on PTB-XL demonstrates that the model not only surpasses established methods but also achieves a new level of performance in multi-label ECG classification. Specifically, our method attains an overall Accuracy of 87.59\% and an AUC of 91.33\%, an improvement of approximately 4\% in Accuracy and 1\% in AUC compared to standard LSTM frameworks.

\subsection{Extended Experiments}

To evaluate the robustness of our xLSTM-ECG method, we conduct additional experiments on the Georgia 12-Lead ECG dataset~\cite{alday2020classification}. Unlike PTB-XL, which contains 5 categories, this dataset consists of 7 categories: NSR, AF, IAVB, LBBB, RBBB, SB, and STach. In our setup, group No.~1 is held out for testing, while the remaining groups (No.~2--11) are used for training. We modify the classification heads illustrated in Figure~\ref{fig:arch} to accommodate the expanded label set and train our model from scratch for 20 epochs.

\begin{figure}[h!]
  \centering
  \begin{subfigure}[b]{0.48\textwidth}
  \centering
  \includegraphics[width=\textwidth]{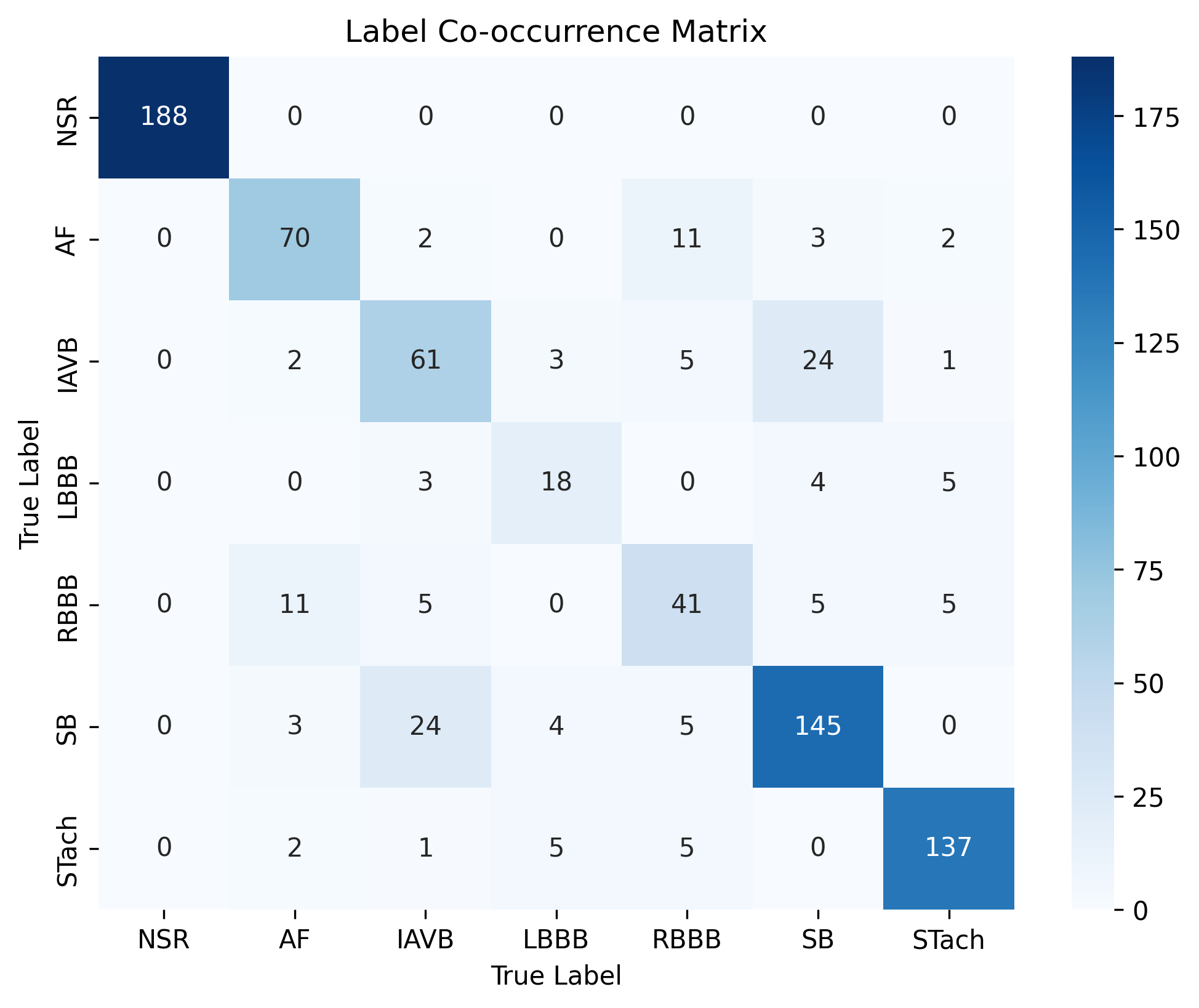}
  \caption{True vs. true label co-occurrence matrix.}
  \end{subfigure}
\hfill
  \begin{subfigure}[b]{0.48\textwidth}
  \centering
  \includegraphics[width=\textwidth]{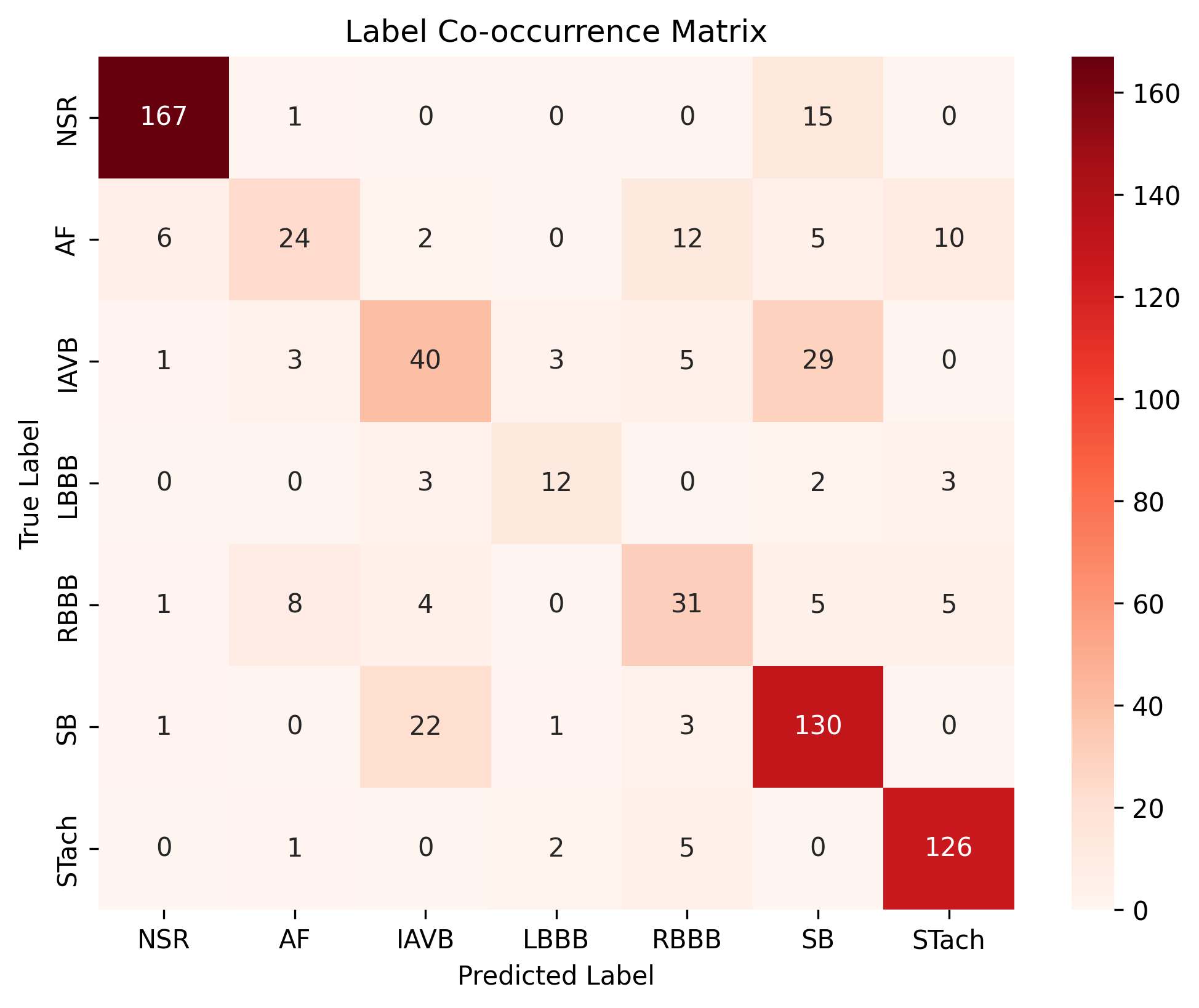}
  \caption{True vs. predicted label co-occurrence matrix.}
  \end{subfigure}
  \caption{Label co-occurrence matrices for multi-class, multi-label classification on Georgia 12-Lead ECG dataset: (a) shows the ground-truth co-occurrence matrix, and (b) shows the predicted co-occurrence matrix.}
  \label{fig:cooccurence_georgia}
\end{figure}

\begin{figure}[ht]
  \centering
  \includegraphics[width=0.98\linewidth]{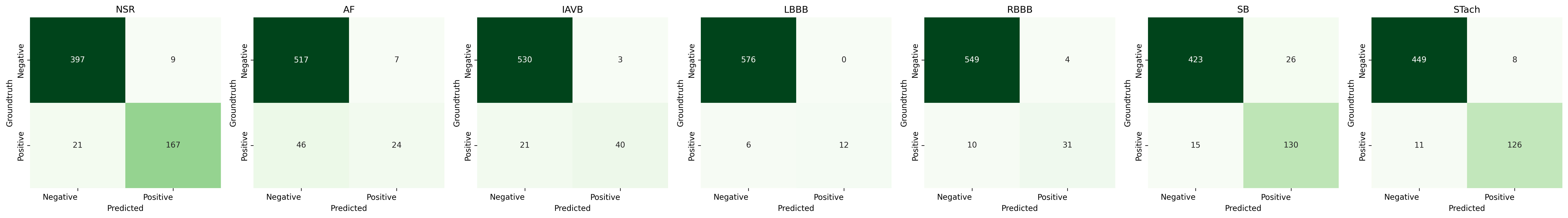}
  \caption{Confusion matrices corresponding to the seven categoriesin Georgia 12-Lead ECG dataset: NSR, AF, IAVB, LBBB, RBBB, SB, and STach.}
  \label{fig:confusion_georgia}
\end{figure}

Figure~\ref{fig:cooccurence_georgia} shows label co-occurrence matrices for both ground-truth and model-predicted labels. Comparing these matrices reveals whether the model accurately captures real-world correlations among classes or systematically overestimates certain co-occurrences. Figure~\ref{fig:confusion_georgia} provides seven confusion matrices—one for each category—detailing correct classifications (diagonal entries) and misclassifications (off-diagonal entries). These visualizations shed light on which classes are most commonly confused and where additional refinement might be necessary to boost clinical reliability.

\begin{table}[ht]
    \caption{Experiments on Georgia 12-Lead ECG dataset with different number of layers $\text{N}_{layers}$.}
    \label{tab:results_georgia}
    \centering
    \small
    \scalebox{0.88}{
    \begin{tabular}{cccccccc}
    \toprule
    \textbf{Method} & \textbf{$\text{N}_{layers}$} & \textbf{Accuracy (\%)} & \textbf{Precision (\%)} & \textbf{Recall (\%)} & \textbf{$\text{F1}_{macro}$ (\%)}  &  \textbf{AUC (\%)} & \textbf{MAP (\%)}\\
    \midrule
    xLSTM-ECG (Proposed) & 2 & 95.02 & 89.97 & 69.80 & 75.29 & \textbf{95.70} & \textbf{86.56}\\
    xLSTM-ECG (Proposed) & 4 & \textbf{95.50}  & \textbf{90.18} & \textbf{73.23} & \textbf{79.59} & 95.44 & 85.99\\
    xLSTM-ECG (Proposed) & 6 & 94.93 & 88.90 & 67.57 & 74.21 & 94.78 & 84.29\\
    \bottomrule
    \end{tabular}
    }
\end{table}

Table~\ref{tab:results_georgia} summarizes our findings under different network depths. With a 2-layer configuration, the model achieves high Accuracy (95.02\%) and Precision (89.97\%), but the Recall (69.80\%) and macro F1 score (75.29\%) lag behind, even though the AUC remains strong at 95.70\%. Increasing the depth to 4 layers yields a more balanced outcome across Accuracy (95.50\%), Precision (90.18\%), Recall (73.23\%), and macro F1 (79.59\%), suggesting that a 4-layer architecture effectively captures both short-term and long-term ECG features. Although the AUC (95.44\%) and MAP (85.99\%) are slightly lower than in the 2-layer case, the overall performance at 4 layers appears optimal for most metrics. Further depth expansion to 6 layers results in a decline in Accuracy, Precision, and Recall, indicating potential overfitting or diminishing returns when adding more layers.

These results highlight the adaptability of our xLSTM-ECG approach, which can be retargeted to new datasets with minimal modification. Despite the more complex label structure of the Georgia 12-Lead ECG dataset, the model consistently demonstrates robust classification performance, reinforcing its suitability for real-world clinical applications where data characteristics may vary considerably across institutions and patient populations.

\section{Conclusion}

By combining the sLSTM and mLSTM modules within the xLSTM framework, our model captures both local and global patterns in 12-lead ECG signals, delivering superior performance on the PTB-XL dataset (87.59\% Accuracy and 91.33\% AUC). This represents an approximate 4\% Accuracy and 1\% AUC improvement over traditional LSTM approaches, highlighting the effectiveness of extended memory structures in modeling inter-lead dependencies. Conditions such as Hypertrophy (HYP) and Myocardial Infarction (MI) still pose challenges; ongoing work involving adaptive attention fusion, multi-scale preprocessing (for example, combining STFT and wavelets), and additional evaluation metrics could enhance the model’s ability to identify subtle or rare classes.

In addition to these performance gains, the xLSTM-ECG model has significant clinical implications. By reducing misdiagnoses of overlapping cardiac conditions, it offers the potential for more precise and timely care. This study also demonstrates how deep learning architectures can be specialized for complex, multi-dimensional biomedical signals, providing a pathway for future research that integrates multiple modalities, such as text or imaging, to advance computational medicine. With proven scalability and robustness using large-scale data, our approach supplies a strong framework for refining ECG analysis pipelines, ultimately aiming to improve patient outcomes and revolutionize diagnostic practices in various clinical settings.

\section*{Acknowledgment}

Beatriu de Pinós del Departament de Recerca i Universitats de la Generalitat de Catalunya (2022 BP 00256). The predoctoral program AGAUR-FI ajuts (2024 FI-3 00065) Joan Oró, which is backed by the Secretariat of Universities and Research of the Department of Research and Universities of the Generalitat of Catalonia, as well as the European Social Plus Fund. European Lighthouse on Safe and Secure AI (ELSA) from the European Union’s Horizon Europe programme under grant agreement No 101070617.

\bibliographystyle{plain}
\bibliography{references}

\end{document}